\documentclass[runningheads]{llncs}
\usepackage[T1]{fontenc}
\usepackage{amsmath}
\usepackage{algorithm}
\usepackage{algpseudocode}

\usepackage{graphicx,verbatim}
\begin{document}
\title{Beyond Manual Annotation: A Human-AI Collaborative Framework for Medical Image Segmentation Using Only “Better or Worse” Expert Feedback}


\titlerunning{Segmentation from Better-or-Worse Feedback}
%

\author{Yizhe Zhang}  
\authorrunning{Y. Zhang}
\institute{Nanjing University of Science and Technology \\
    \email{zhangyizhe@njust.edu.cn}}

\maketitle              
\begin{abstract}
Manual annotation of medical images is a labor-intensive and time-consuming process, posing a significant bottleneck in the development and deployment of robust medical imaging AI systems. This paper introduces a novel hands-free Human-AI collaborative framework for medical image segmentation that substantially reduces the annotation burden by eliminating the need for explicit manual labeling. The core innovation lies in a preference learning paradigm, where human experts provide minimal, intuitive feedback—simply indicating whether an AI-generated segmentation is better or worse than a previous version. The framework comprises four key components: (1) an adaptable foundation model (FM) for feature extraction, (2) label propagation based on feature similarity, (3) a clicking agent that learns from human better-or-worse feedback to decide where to click and with which label, and (4) a multi-round segmentation learning procedure that trains a state-of-the-art segmentation network using pseudo-labels generated by the clicking agent and FM-based label propagation. Experiments on three public datasets demonstrate that the proposed approach achieves competitive segmentation performance using only binary preference feedback—without requiring experts to directly manually annotate the images. 


\end{abstract}
\section{Introduction}

The delineation of anatomical structures and pathological regions through image segmentation is a cornerstone of modern medical image analysis, critical for clinical diagnostics, treatment planning, and longitudinal studies \cite{litjens2017survey,konwer2025enhancing}. The advent of deep learning, particularly convolutional neural networks (e.g.,~\cite{ronneberger2015u},~\cite{isensee2021nnu}) and transformer-based architectures (e.g.,~\cite{dosovitskiy2021image}), has led to state-of-the-art performance in numerous medical segmentation tasks. However, the success of these supervised learning models is predicated on the availability of large-scale, high-quality, pixel-level annotations. The generation of these annotations is a significant bottleneck, being notoriously slow, expensive, and requiring extensive domain expertise \cite{liao2024modeling,cheplygina2019not}. This reliance on meticulous manual labeling hinders the scalability and rapid deployment of AI-powered segmentation tools in diverse clinical settings.

\begin{figure}[t]
\centering
\includegraphics[width=1.0\textwidth]{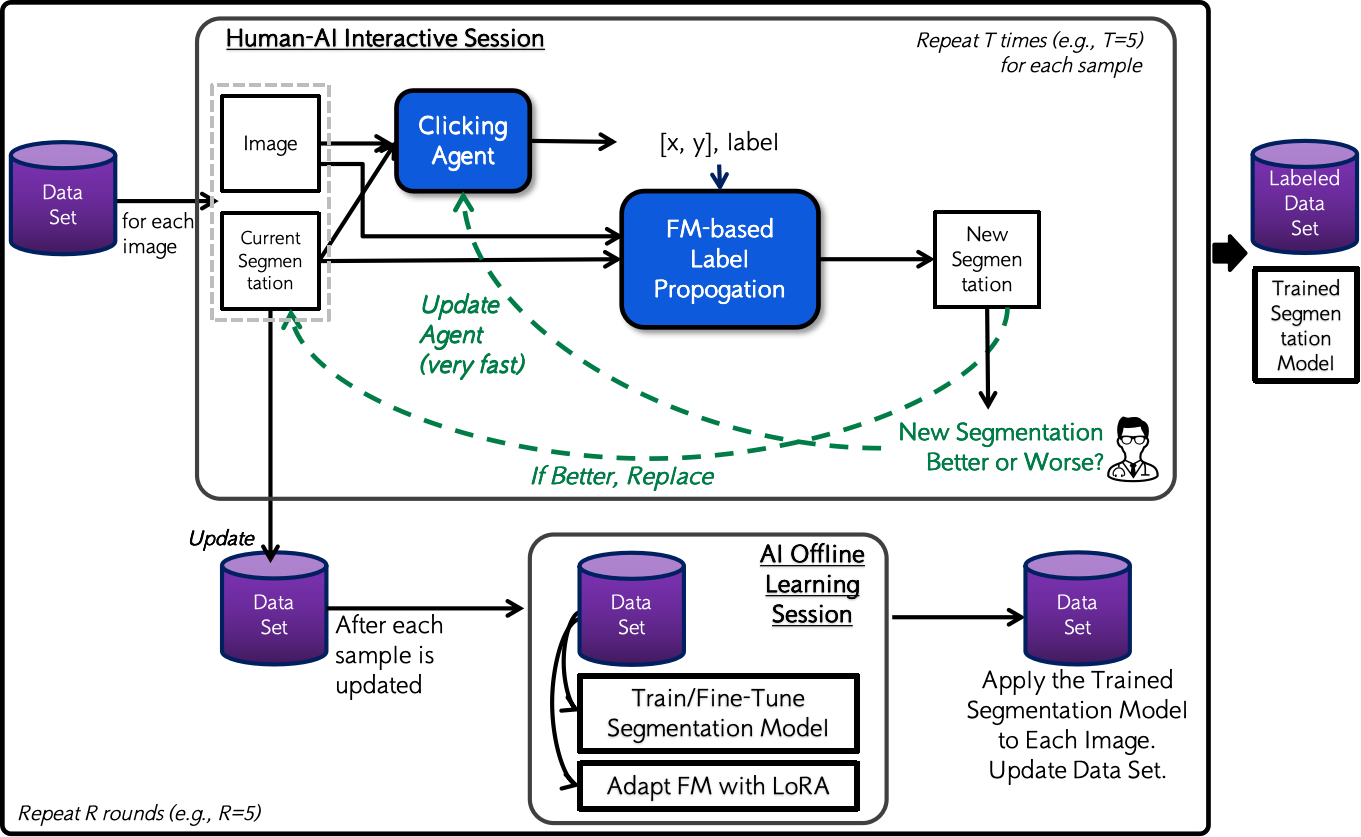}
\caption{Overview of the proposed framework.} \label{fig1}
\end{figure}

To mitigate this annotation burden, the community has explored various weakly supervised and interactive segmentation paradigms. These methods aim to reduce the annotation effort by leveraging simpler forms of human guidance, such as bounding boxes, scribbles, or points \cite{wang2018interactive,can2018learning}. More recently, the emergence of large-scale, pre-trained foundation models, such as the Segment Anything Model (SAM), has marked a significant shift \cite{kirillov2023segment}. These models, trained on vast datasets of natural images, exhibit remarkable zero-shot and few-shot generalization capabilities, enabling them to segment objects based on minimal user prompts like clicks or boxes. Adaptations of these models for the medical domain, such as MedSAM \cite{ma2024segment} and SAM-Med2D \cite{cheng2023sam}, have further demonstrated their potential to reduce annotation time.

Despite these advancements, existing interactive approaches, including those based on foundation models, still require the user to \textbf{manually} provide explicit and spatially precise inputs. The expert must still identify correct and incorrect regions, a task that, while simpler than full manual segmentation, can still be cognitively demanding and time-consuming, especially for complex 3D anatomies or subtle pathologies, especially for complex 3D anatomies or subtle pathologies~\cite{liao2020iteratively}. Furthermore, these methods often necessitate multiple rounds of corrective interactions to refine the segmentation to a clinically acceptable standard.

A more intuitive and less burdensome form of expert feedback would be comparative. Often, a clinician can readily judge whether one segmentation output is ``better'' or ``worse'' than another, even if they cannot, or do not have the time to, pinpoint the exact locations of all errors. This type of preference-based feedback is natural to human decision-making and has the potential to significantly streamline the human-AI interaction process. Recent work has begun to explore the use of preference learning for model optimization, for instance, by using direct preference optimization to refine foundation models with simulated annotator rankings \cite{konwer2025enhancing}.

In this paper, we introduce a novel human-AI collaborative framework that moves beyond traditional explicit annotation and harnesses the power of simple, binary preference feedback. Our method entirely bypasses the need for manually providing pixel-level ground truth, instead learning to produce accurate segmentations from an expert's comparative judgments and feedback. The core of our framework is a learning-based clicking agent that, guided by ``better or worse'' signals from a human expert, intelligently samples points to generate pseudo-labels. These pseudo-labels, propagated in the rich feature space of a foundation model, are then used to train a pre-trained, state-of-the-art segmentation network. Our main contributions are:
\begin{itemize}
    \item A novel human-AI collaborative segmentation framework that learns from minimal, non-local, binary preference feedback (i.e., ``better or worse''), eliminating the need for any manual pixel-level and/or region-level annotation.
    \item A learning-based clicking agent that interprets the expert's preference feedback to strategically decide where to sample points and with which labels to generate high-quality pseudo-annotations.
    \item The integration of a foundation model for robust feature extraction and label propagation, enabling the generation of dense pseudo-labels from the sparse points provided by the clicking agent.
    \item A demonstration on multiple public medical imaging datasets that our preference-based approach can achieve competitive segmentation performance.
\end{itemize}
Our work represents a significant step toward more efficient, scalable, and user-friendly AI systems for medical image segmentation, enabling high-quality results even when detailed manual annotations are costly or impractical to obtain. A high-level overview of the proposed method is illustrated in Fig.~\ref{fig1}.

\section{Method}
Our proposed Human-AI collaborative framework learns to segment medical images from simple ``better or worse'' feedback, obviating the need for manual pixel-level annotations. The framework operates in an iterative, multi-round process, where each round consists of two main stages: (1) an interactive pseudo-labeling stage where a clicking agent, guided by expert preference, generates sparse annotations that are propagated into dense pseudo-masks; and (2) a segmentation model training stage where a state-of-the-art network is trained on these pseudo-labels.

\subsection{Adaptable Foundation Model for Feature Extraction}

The cornerstone of our label propagation mechanism is a large, pre-trained foundation model. We employ DINOv2~\cite{kirillov2023segment} with a Vision Transformer (ViT) backbone, which provides a rich, patch-level feature space that captures fine-grained semantic and spatial relationships within the image. Given an input image, the model outputs a sequence of normalized feature vectors $\{f_i\}_{i=1}^{N}$, where $N$ is the number of patches.

To tailor the generic features of DINOv2 to the specific medical domain of the target dataset, we make the model adaptable. We employ Low-Rank Adaptation (LoRA)~\cite{hu2022lora}, which introduces trainable, low-rank matrices into the query and value projections of the transformer's self-attention layers. This allows us to efficiently fine-tune the feature extractor using a contrastive triplet loss on the pseudo-labeled data gathered during the interactive phase. This adaptation sharpens the feature space, improving the model's ability to distinguish between foreground and background tissue based on the evolving pseudo-annotations.

\subsection{Clicking Agent and Label Propagation}

\subsubsection{Preference-Based Clicking Agent}
We formalize the task of identifying informative correction locations as a Reinforcement Learning (RL) problem. We design a lightweight ``clicking agent'' whose goal is to learn an optimal policy for selecting click coordinates.
\begin{itemize}
\item{State.} The state $s_t$ at step $t$ is a multi-channel tensor representing the agent's current knowledge. It comprises the resized input image concatenated with the current predicted segmentation mask. This provides the agent with both the original image context and the current state of the segmentation it needs to improve.

\item{Action.} The agent's action $a_t$ is the selection of a single pixel coordinate $(y, x)$ on which to place a corrective click. The agent's policy is modeled by a small U-Net, which outputs a logit map over the input image space. An action is sampled from the resulting softmax probability distribution, with a temperature parameter to control exploration.

\item{Reward.} The reward signal $r_t$ directly models the ``better or worse'' feedback. After the agent selects a click location, we use it to update the segmentation via label propagation. The resulting new segmentation mask is compared to the mask before the click. The reward is binary: $+1$ if expert consider the result gets better and $-1$ if it does not (a ``worse'' outcome). The agent's policy is updated using the REINFORCE algorithm~\cite{williams1992simple}, a policy gradient method that adjusts its parameters to favor actions that yield higher rewards and improve segmentation.
\end{itemize}

\subsubsection{DINO-based Label Propagation}\label{sec:dino-prop}
A single click provides only sparse information. To generate a dense pseudo-mask for training the segmentation model, we propagate the click's label based on feature similarity. When a click is placed at coordinate $(y,x)$ with label $l \in \{\text{foreground}, \text{background}\}$, we identify the corresponding feature vector $f_{click}$ from our adapted DINOv2 model. We then compute the cosine similarity between $f_{click}$ and all other patch features $\{f_i\}$ in the image. All patches whose feature similarity to $f_{click}$ exceeds a predefined threshold (e.g., 0.8) are assigned the label $l$. This process converts a few sparse clicks into a dense pseudo-label map.

\subsection{Multi-Round Segmentation Model Training}
The pseudo-labels generated from the interactive process across the dataset form the training set for a dedicated segmentation network. At the end of each round, a segmentation model is trained or fine-tuned on the collection of newly generated pseudo-masks. The refined model from the current round then serves as the baseline model for the next round, providing progressively better initial segmentations for the agent to improve upon. This iterative refinement allows the model to learn complex anatomical features from simple preference feedback, bootstrapping its performance over several rounds. To further enhance performance, we optionally filter the generated pseudo-labels, using only the top-K percent high quality pseudo-labels to train the segmentation model. The complete procedure of the proposed method is outlined in Algorithm 1.

\begin{algorithm}[t]
\scriptsize
\caption{Human-AI Collaborative Annotation using Preference Feedback}
\label{alg:framework}
\begin{algorithmic}[1]
\State \textbf{Initialize:} Segmentation model $S_{seg}$, Clicking Agent $A_{click}$, Adaptable Foundation Model $\mathcal{F}_{adapt}$ (DINOv2+LoRA).
\State \textbf{Input:} Dataset $\mathcal{D}$, number of rounds $R=5$, interaction steps per image $T=5$, similarity threshold $\tau = 0.8$.

\For{round $r=1$ to $R$}
    \State $\mathcal{P}_r \gets \emptyset$ \Comment{Initialize set of pseudo-labels for the current round}
    \For{each image $I \in \mathcal{D}$}
        \State Extract patch features $\{f_i\} \gets \mathcal{F}_{adapt}(I)$
        \State $M_{current} \gets S_{seg}(I)$ \Comment{Get initial prediction from the main segmentation model}
        \For{step $t=1$ to $T$}
            \State Define state $s_t \gets \text{concat}(I, M_{current})$
            \State Sample click action $a_t \sim A_{click}(s_t)$ \Comment{Agent selects a click location}
            \State $M_{new} \gets \text{PropagateLabel}(a_t, \{f_i\}, \tau)$ \Comment{Generate mask via DINO-based propagation}
            \State $r_t \gets \text{GetExpertPreference}(M_{new}, M_{current})$ \Comment{Reward is $+1$ if better, $-1$ if worse}
            \State Update $A_{click}$'s policy using $(s_t, a_t, r_t)$ via REINFORCE algorithm.
            \If{$r_t = +1$}
                \State $M_{current} \gets M_{new}$ \Comment{Update mask only on improvement}
            \EndIf
        \EndFor
        \State Add final pseudo-mask $M_{current}$ to $\mathcal{P}_r$.
    \EndFor
    
    \State $\mathcal{P}_{r} \gets \text{FilterTopK}(\mathcal{P}_r)$ \Comment{Optional: Select highest-quality pseudo-labels}
    \State Fine-tune $S_{seg}$ on $\mathcal{P}_{r}$.
    \State Adapt $\mathcal{F}_{adapt}$ on $\mathcal{P}_{r}$ using a contrastive triplet loss.
\EndFor
\State \textbf{Return:} Fully trained segmentation model $S_{seg}$.
\end{algorithmic}
\end{algorithm}

\section{Experiments}

We conducted a series of experiments to validate our proposed framework's ability to generate high-quality pseudo-labels and train an effective segmentation model using only preference-based feedback (``better or worse" feedback). Our evaluation is performed on three distinct and publicly available medical image segmentation datasets, covering different modalities and anatomical targets.
\noindent\textbf{Breast Ultrasound Segmentation.} To evaluate our method on ultrasound imagery, we use the Breast Ultrasound Images (BUSI) dataset \cite{al2020dataset}. This dataset contains 780 ultrasound scans categorized as normal, benign, and malignant, each with a corresponding ground truth mask. As the original release does not specify a train/test split, we utilize the complete dataset for our experiments.

\noindent\textbf{Skin Lesion Segmentation.} For dermoscopic image analysis, we use the training set from the ISIC 2018 Challenge~\cite{tschandl2018ham10000}. This set contains 2594 images of skin lesions with corresponding binary segmentation masks. Similar to the polyp dataset, we use all provided training images to validate the proposed method.

\noindent\textbf{Polyp Segmentation.} 
We use 1,450 colonoscopy images collected from two polyp segmentation datasets: {Kvasir-SEG}~\cite{jha2019kvasir} and {CVC-ClinicDB}~\cite{bernal2015wm}. All images are accompanied by high-quality manual segmentation masks. The combined dataset captures a wide range of polyp sizes, shapes, and appearances.

\subsection{Experimental Setup}
We use the provided ground truth masks in these datasets only as an \textbf{oracle} to simulate the expert's ``better or worse" feedback, not as supervised training masks for the segmentation model. To simulate expert feedback, we use the ground truth mask as an oracle. We compute the Dice Similarity Coefficient (DSC) between the oracle and the segmentation mask both before and after the agent's click. If the DSC increases, the outcome is labeled ``better'' (reward = +1). If not, it is labeled ``worse'' (reward = -1). This binary signal is the sole feedback used to train the clicking agent. We set the number of learning rounds to 5 across all experiments. In each round, every image receives 5 clicks from the clicking agent. A PVT-based~\cite{wang2022pvt} segmentation model, i.e., HSNet~\cite{zhang2022hsnet}, is employed as the main segmentation network for all experiments.

\subsection{Results and Analyses}

\begin{figure}[t]
\centering
\includegraphics[width=1.0\textwidth]{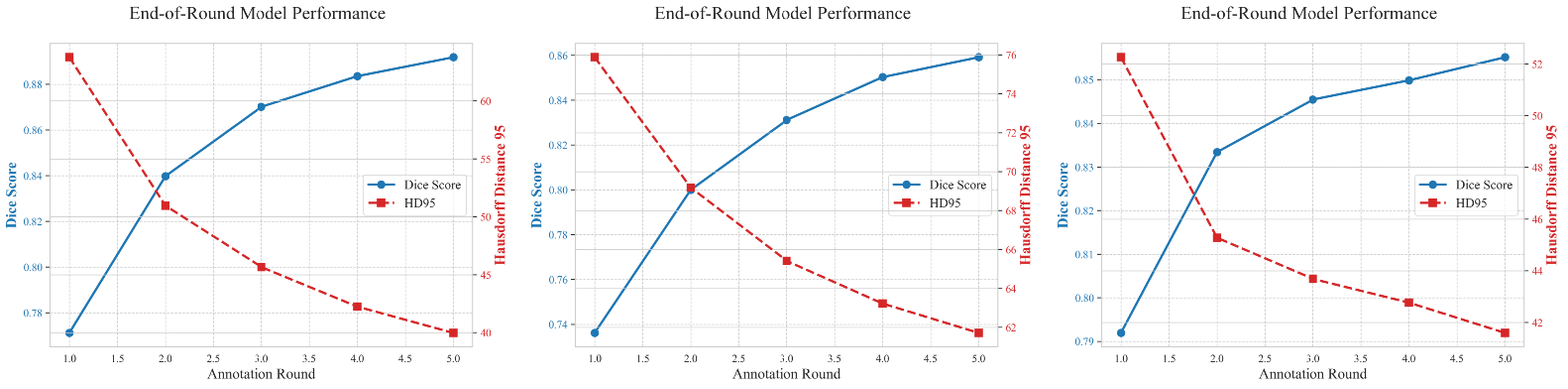}
\caption{Segmentation performance across annotation rounds, showing progressive improvement with increasing Dice scores and decreasing HD95 on polyp (left), skin lesion (middle), and ultrasound (right) datasets.} \label{fig2}
\end{figure}

\begin{figure}[t]
\centering
\includegraphics[width=1.0\textwidth]{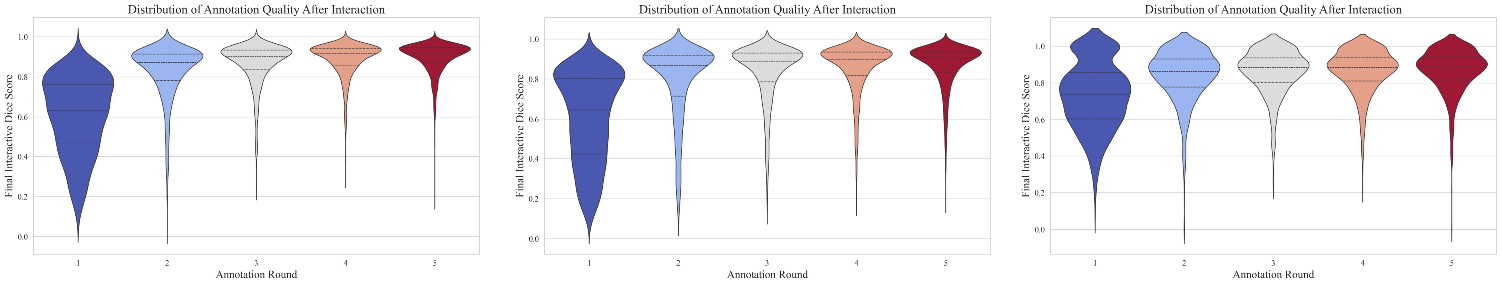}
\caption{Violin plots showing the distribution of final interactive Dice scores across annotation rounds, highlighting improved annotation quality and consistency over time. Results are shown for polyp (left), skin lesion (middle), and ultrasound (right) datasets.} \label{fig3}
\end{figure}


\begin{figure}[t]
\centering
\includegraphics[width=0.8\textwidth]{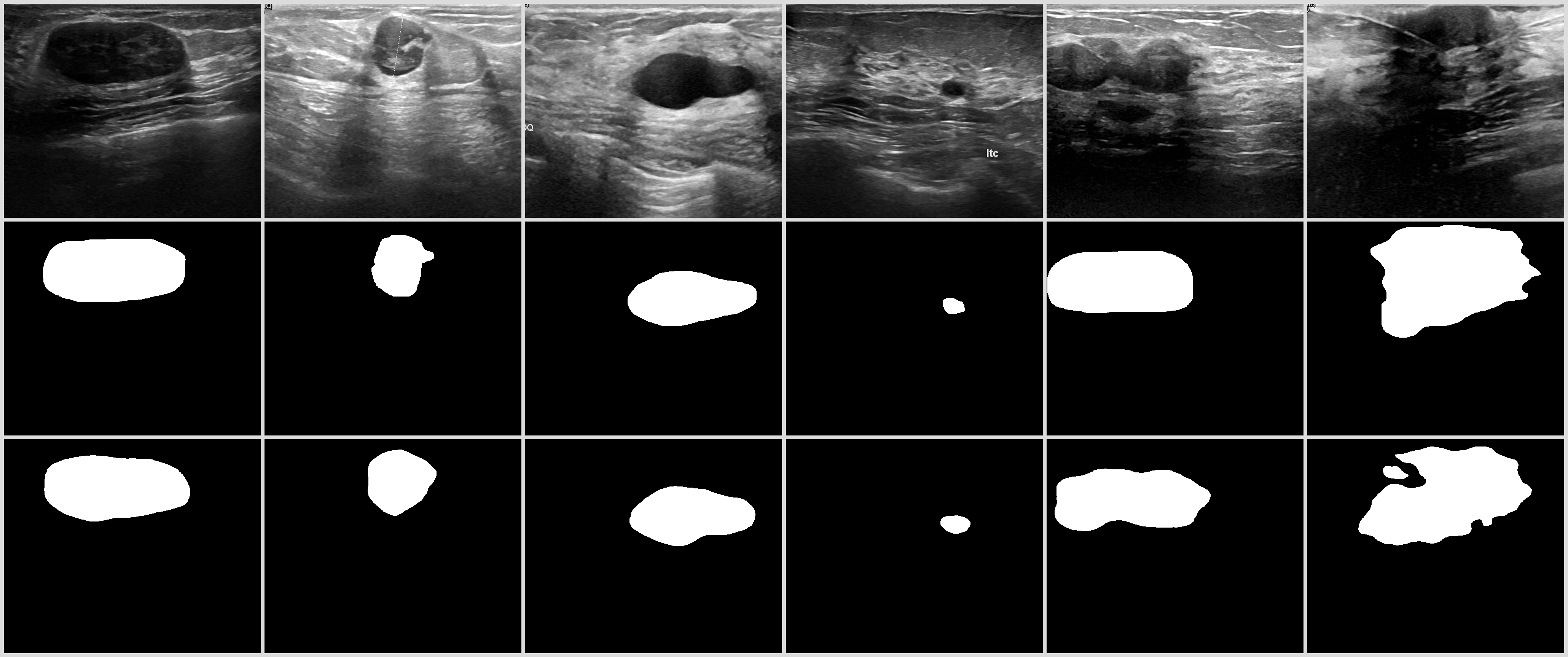}
\caption{Segmentation examples on the BUSI dataset. Top: input ultrasound image; Middle: ground truth; Bottom: prediction by our method.} \label{fig5}
\end{figure}

\begin{figure}[t]
\centering
\includegraphics[width=0.8\textwidth]{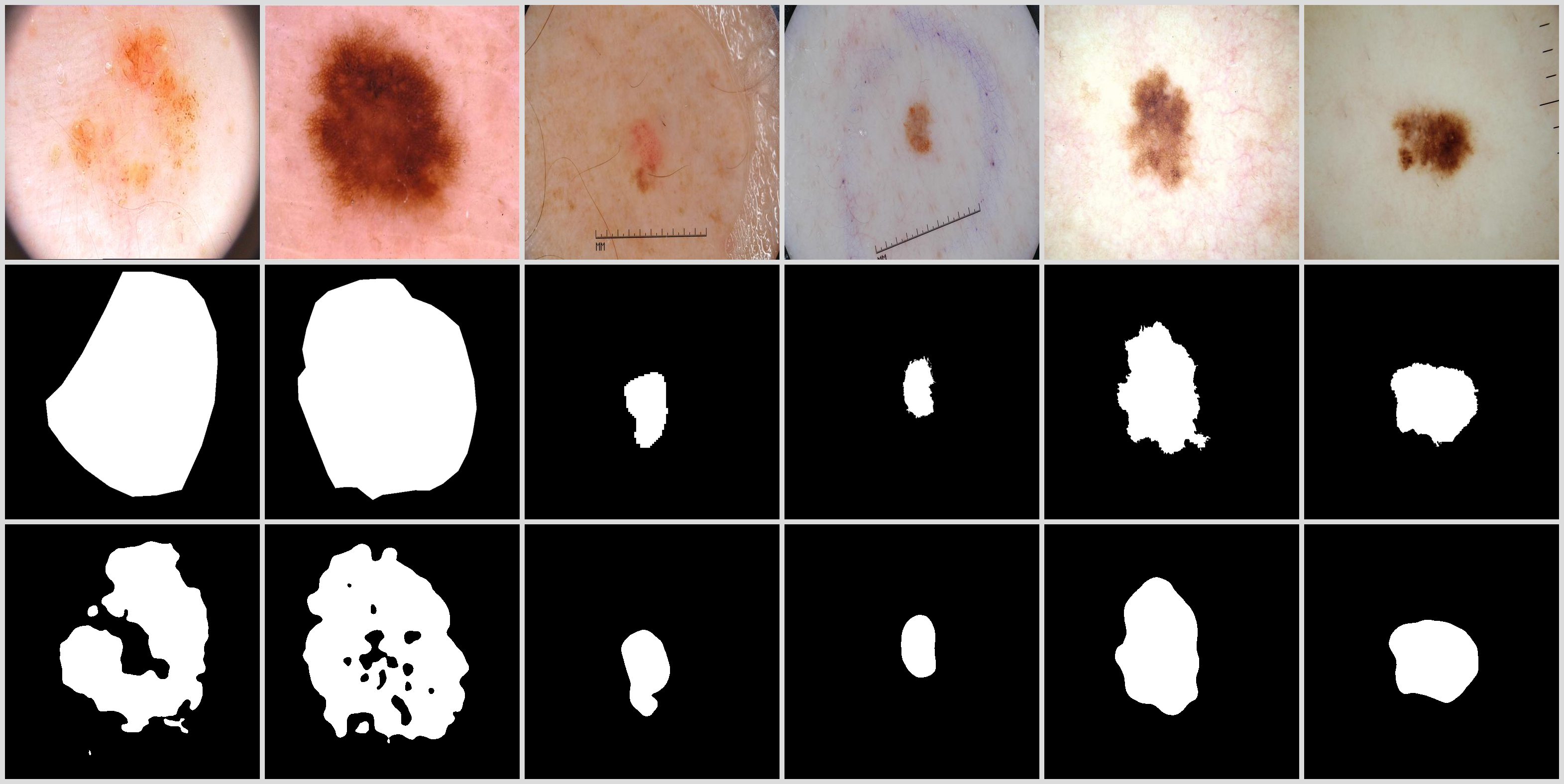}
\caption{Segmentation examples on the skin lesion dataset. Top: input dermoscopy image; Middle: ground truth mask; Bottom: segmentation predicted by our method.} \label{fig6}
\end{figure}

In Fig.~\ref{fig2}, we present the end-of-round segmentation performance across three diverse medical image segmentation tasks: ultrasound image segmentation (left), skin lesion segmentation (middle), and polyp segmentation (right). As the annotation rounds progress, we observe a consistent improvement in segmentation quality across all datasets, evidenced by increasing Dice scores and decreasing Hausdorff Distance at 95th percentile (HD95). This trend demonstrates the effectiveness of our proposed human-AI collaborative framework in incrementally enhancing model performance through minimal expert feedback. Interestingly, we observed that segmentation performance on ultrasound images was consistently lower compared to other modalities. This suggests the need for more domain-specific foundation models tailored to ultrasound imaging. In our current framework, the feature extractor is based on DINO-v2, a general-purpose foundation model primarily trained on natural scene images, which may limit its effectiveness in capturing ultrasound-specific characteristics.

\begin{figure}[t]
\centering
\includegraphics[width=0.8\textwidth]{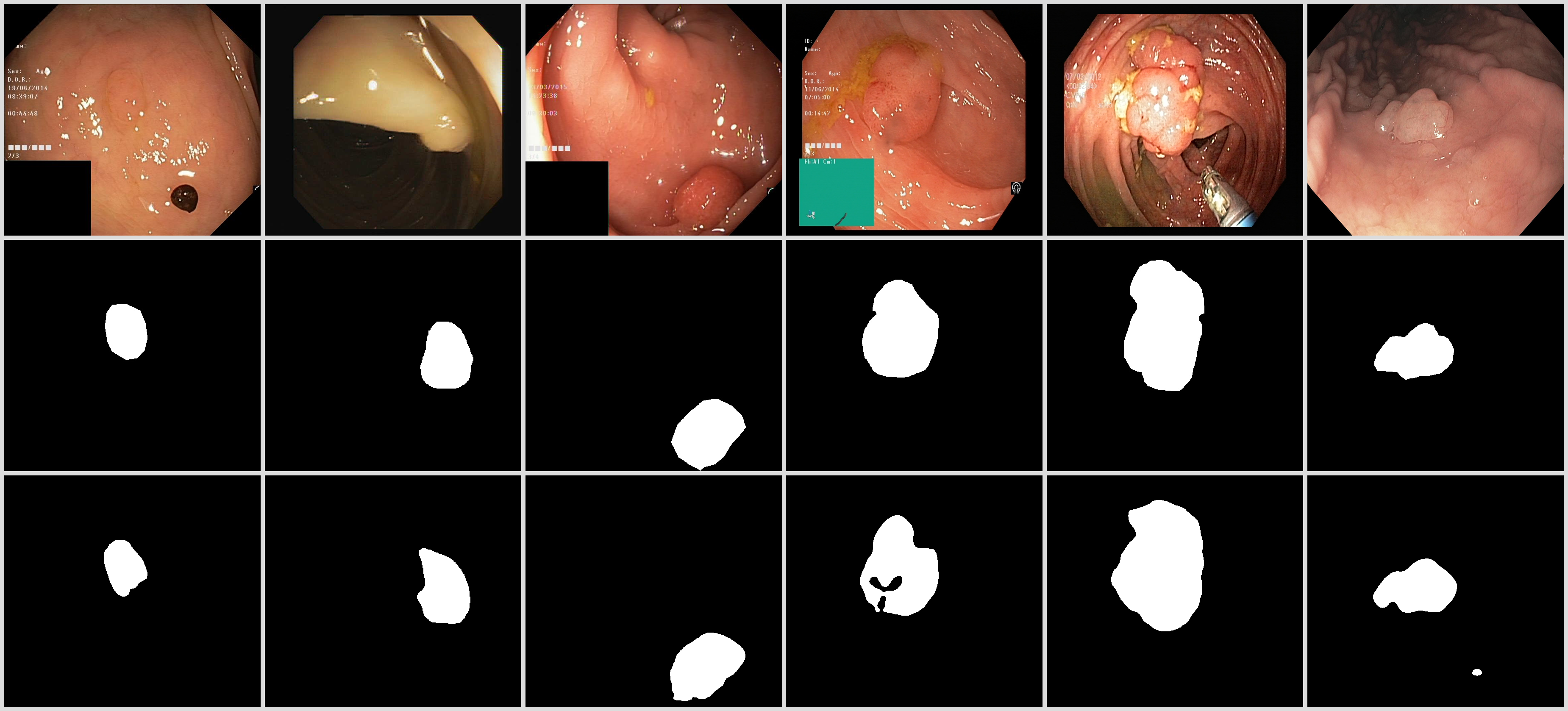}
\caption{Segmentation examples on the polyp dataset. Top: input endoscopic image; Middle: ground truth mask; Bottom: segmentation predicted by our method.} \label{fig7}
\end{figure}

Fig.~\ref{fig3} illustrates the distribution of segmentation quality, measured by the Dice score, after each round of interaction across the three datasets: ultrasound (left), skin lesion (middle), and polyp (right). Across all tasks, we observe a clear shift in the distribution toward higher Dice scores as annotation rounds progress, indicating that both the average and consistency of segmentation quality improve with our method. Notably, the initial annotation round exhibits a wide and often bimodal distribution, reflecting substantial variability in interaction effectiveness. However, from the second round onward, the distributions become increasingly concentrated around higher Dice scores (often exceeding 0.8), demonstrating that our clicking agent, guided by simple ``better or worse'' feedback, learns to produce more accurate and reliable annotations over time. Fig.~\ref{fig5}, Fig.~\ref{fig6}, and Fig.~\ref{fig7} present visualizations of the segmentation results. Although our method performs well overall, certain cases still highlight areas for potential improvement.


\section{Limitations}
Despite promising results, our framework has limitations. First, we rely on a simulated oracle that compares Dice scores to generate “better or worse” feedback. While this enables controlled evaluation, it oversimplifies real-world expert judgments, which are often subjective and prioritize clinical relevance over global metrics. Human feedback also introduces variability and cognitive bias not captured by our oracle. Second, the learning efficiency of the clicking agent remains a challenge. The REINFORCE-based training suffers from high variance and sparse binary rewards, which may delay convergence and require many interactions before the agent becomes effective.

\section{Conclusion}
This paper tackles the persistent bottleneck of manual annotation in medical image segmentation by introducing a novel human-AI collaborative framework that learns entirely from ``better or worse'' expert feedback. We have demonstrated that by combining a learning-based clicking agent, a pre-trained foundation model for feature extraction and label propagation, and a multi-round training strategy, it is possible to train a high-performance segmentation network without requiring any direct manual annotation from users. Our experiments across polyp, skin lesion, and breast ultrasound datasets validate that this minimal form of feedback is sufficient to achieve competitive segmentation accuracy. This work marks a promising step towards creating more intuitive, efficient, and scalable AI systems for medical imaging, potentially shifting the human's role from a tedious annotator to a high-level critic. 


\bibliographystyle{splncs04} 
\bibliography{example}

\begin{thebibliography}{10}
\providecommand{\url}[1]{\texttt{#1}}
\providecommand{\urlprefix}{URL }
\providecommand{\doi}[1]{https://doi.org/#1}

\bibitem{al2020dataset}
Al-Dhabyani, W., Gomaa, M., Khaled, H., Fahmy, A.: Dataset of breast ultrasound images. Data in brief  \textbf{28},  104863 (2020)

\bibitem{bernal2015wm}
Bernal, J., S{\'a}nchez, F.J., Fern{\'a}ndez-Esparrach, G., Gil, D., Rodr{\'\i}guez, C., Vilari{\~n}o, F.: Wm-dova maps for accurate polyp highlighting in colonoscopy: Validation vs. saliency maps from physicians. Computerized medical imaging and graphics  \textbf{43},  99--111 (2015)

\bibitem{can2018learning}
Can, Y.B., Chaitanya, K., Mustafa, B., Koch, L.M., Konukoglu, E., Baumgartner, C.F.: Learning to segment medical images with scribble-supervision alone. In: Deep Learning in Medical Image Analysis and Multimodal Learning for Clinical Decision Support: 4th International Workshop, DLMIA 2018, and 8th International Workshop, ML-CDS 2018, Held in Conjunction with MICCAI 2018, Granada, Spain, September 20, 2018, Proceedings 4. pp. 236--244. Springer (2018)

\bibitem{cheng2023sam}
Cheng, J., Ye, J., Deng, Z., Chen, J., Li, T., Wang, H., Su, Y., Huang, Z., Chen, J., Jiang, L., et~al.: Sam-med2d. arXiv preprint arXiv:2308.16184  (2023)

\bibitem{cheplygina2019not}
Cheplygina, V., de~Bruijne, M., Pluim, J.P.: Not-so-supervised: a survey of semi-supervised, multi-instance, and transfer learning in medical image analysis. Medical image analysis  \textbf{54},  280--296 (2019)

\bibitem{dosovitskiy2021image}
Dosovitskiy, A., Beyer, L., Kolesnikov, A., Weissenborn, D., Zhai, X., Unterthiner, T., Dehghani, M., Minderer, M., Heigold, G., Gelly, S., et~al.: An image is worth 16x16 words: Transformers for image recognition at scale. In: International Conference on Learning Representations (2021)

\bibitem{hu2022lora}
Hu, E.J., Shen, Y., Wallis, P., Allen-Zhu, Z., Li, Y., Wang, S., Wang, L., Chen, W., et~al.: Lora: Low-rank adaptation of large language models. ICLR  \textbf{1}(2), ~3 (2022)

\bibitem{isensee2021nnu}
Isensee, F., Jaeger, P.F., Kohl, S.A., Petersen, J., Maier-Hein, K.H.: nnu-net: a self-configuring method for deep learning-based biomedical image segmentation. Nature methods  \textbf{18}(2),  203--211 (2021)

\bibitem{jha2019kvasir}
Jha, D., Smedsrud, P.H., Riegler, M.A., Halvorsen, P., De~Lange, T., Johansen, D., Johansen, H.D.: Kvasir-seg: A segmented polyp dataset. In: International conference on multimedia modeling. pp. 451--462. Springer (2019)

\bibitem{kirillov2023segment}
Kirillov, A., Mintun, E., Ravi, N., Mao, H., Rolland, C., Gustafson, L., Xiao, T., Whitehead, S., Berg, A.C., Lo, W.Y., et~al.: Segment anything. In: Proceedings of the IEEE/CVF International Conference on Computer Vision. pp. 3879--3893 (2023)

\bibitem{konwer2025enhancing}
Konwer, A., Yang, Z., Bas, E., Xiao, C., Prasanna, P., Bhatia, P., Kass-Hout, T.: Enhancing sam with efficient prompting and preference optimization for semi-supervised medical image segmentation. In: Proceedings of the Computer Vision and Pattern Recognition Conference. pp. 20990--21000 (2025)

\bibitem{liao2020iteratively}
Liao, X., Li, W., Xu, Q., Wang, X., Jin, B., Zhang, X., Wang, Y., Zhang, Y.: Iteratively-refined interactive 3d medical image segmentation with multi-agent reinforcement learning. In: Proceedings of the IEEE/CVF conference on computer vision and pattern recognition. pp. 9394--9402 (2020)

\bibitem{liao2024modeling}
Liao, Z., Hu, S., Xie, Y., Xia, Y.: Modeling annotator preference and stochastic annotation error for medical image segmentation. Medical Image Analysis  \textbf{92},  103028 (2024)

\bibitem{litjens2017survey}
Litjens, G., Kooi, T., Bejnordi, B.E., Setio, A.A.A., Ciompi, F., Ghafoorian, M., Van Der~Laak, J.A., Van~Ginneken, B., S{\'a}nchez, C.I.: A survey on deep learning in medical image analysis. Medical image analysis  \textbf{42},  60--88 (2017)

\bibitem{ma2024segment}
Ma, J., He, Y., Li, F., Han, L., You, C., Wang, B.: Segment anything in medical images. Nature Communications  \textbf{15}(1), ~654 (2024)

\bibitem{ronneberger2015u}
Ronneberger, O., Fischer, P., Brox, T.: U-net: Convolutional networks for biomedical image segmentation. In: International Conference on Medical image computing and computer-assisted intervention. pp. 234--241. Springer (2015)

\bibitem{tschandl2018ham10000}
Tschandl, P., Rosendahl, C., Kittler, H.: The ham10000 dataset, a large collection of multi-source dermatoscopic images of common pigmented skin lesions. Scientific data  \textbf{5}(1), ~1--9 (2018)

\bibitem{wang2018interactive}
Wang, G., Li, W., Zuluaga, M.A., Pratt, R., Patel, P.A., Aertsen, M., Doel, T., David, A.L., Deprest, J., Ourselin, S., et~al.: Interactive medical image segmentation using deep learning with image-specific fine tuning. IEEE transactions on medical imaging  \textbf{37}(7),  1562--1573 (2018)

\bibitem{wang2022pvt}
Wang, W., Xie, E., Li, X., Fan, D.P., Song, K., Liang, D., Lu, T., Luo, P., Shao, L.: Pvt v2: Improved baselines with pyramid vision transformer. Computational visual media  \textbf{8}(3),  415--424 (2022)

\bibitem{williams1992simple}
Williams, R.J.: Simple statistical gradient-following algorithms for connectionist reinforcement learning. Machine learning  \textbf{8},  229--256 (1992)

\bibitem{zhang2022hsnet}
Zhang, W., Fu, C., Zheng, Y., Zhang, F., Zhao, Y., Sham, C.W.: Hsnet: A hybrid semantic network for polyp segmentation. Computers in biology and medicine  \textbf{150},  106173 (2022)

\end{thebibliography}

\end{document}